# Dynamic causal modelling of phase-amplitude interactions

Erik D. Fagerholm[1,*], Rosalyn J. Moran[1], Inês R. Violante[2], Robert Leech[1,§], Karl J. Friston[3,§]

[1] Centre for Neuroimaging Sciences, Department of Neuroimaging, IoPPN, King's College London
[2] School of Psychology, Faculty of Health and Medical Sciences, University of Surrey
[3] Wellcome Trust Centre for Neuroimaging, Institute of Neurology, University College London

§ These authors contributed equally to this work

* Corresponding author: erik.fagerholm@kcl.ac.uk

**Abstract**

Models of coupled oscillators are used to describe a wide variety of phenomena in neuroimaging. These models typically rest on the premise that oscillator dynamics do not evolve beyond their respective limit cycles, and hence that interactions can be described purely in terms of phase differences. Whilst mathematically convenient, the restrictive nature of phase-only models can limit their explanatory power. We therefore propose a generalisation of dynamic causal modelling that incorporates both phase and amplitude. This allows for the separate quantifications of phase and amplitude contributions to the connectivity between neural regions. We establish, using model-generated data and simulations of coupled pendula, that phase-only models perform well only under weak coupling conditions. We also show that, despite their higher complexity, phase-amplitude models can describe strongly coupled systems more effectively than their phase-only counterparts. We relate our findings to four metrics commonly used in neuroimaging: the Kuramoto order parameter, cross-correlation, phase-lag index, and spectral entropy. We find that, with the exception of spectral entropy, the phase-amplitude model is able to capture all metrics more effectively than the phase-only model. We then demonstrate, using local field potential recordings in rodents and functional magnetic resonance imaging in macaque monkeys, that amplitudes in oscillator models play an important role in describing neural dynamics in anaesthetised brain states.

**Introduction**

Oscillations are observed across a variety of natural systems. In the context of the brain, oscillations may facilitate information exchange (Breakspear et al., 2010; Mejias et al., 2016; Wildegger et al., 2017) at large scales (Buice and Cowan, 2009; Carr, 1981; Haken, 1983) compared with spiking activity at the level of individual neurons. At present, the most compelling evidence for the role of oscillations in human brain function is obtained by perturbing ongoing oscillatory activity, e.g. in the context of Parkinson's disease (Brittain et al., 2013) and sleep studies (Marshall et al., 2006; Ngo et al., 2013). Models of coupled oscillators continue to be essential in describing a variety of neuronal, cognitive and pathogenic settings (Breakspear et al., 2010).

Oscillator models in neuroscience often rest on the assumption that oscillator dynamics are restricted to their respective limit cycles within a narrow torus in phase space (Fig. 1A), defining the boundary of allowable states (Ermentrout and Terman, 2010).

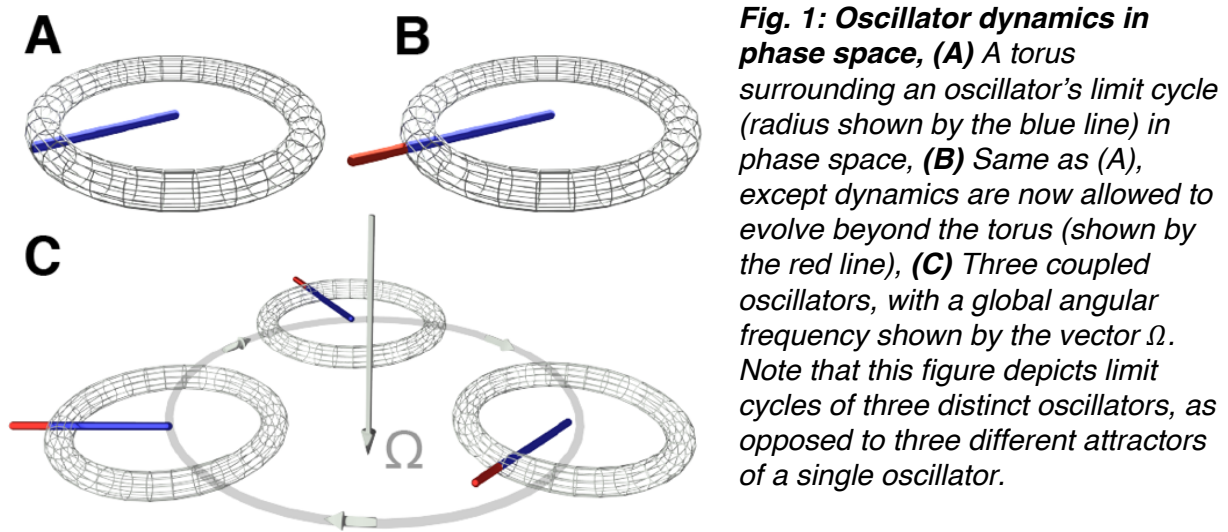

*Fig. 1: Oscillator dynamics in phase space, (A)* A torus surrounding an oscillator's limit cycle (radius shown by the blue line) in phase space, *(B)* Same as (A), except dynamics are now allowed to evolve beyond the torus (shown by the red line), *(C)* Three coupled oscillators, with a global angular frequency shown by the vector $\Omega$. Note that this figure depicts limit cycles of three distinct oscillators, as opposed to three different attractors of a single oscillator.

Note that the term 'phase space' - as used in Fig. 1 - should not be confused with the term 'phase' as used in the context of the instantaneous representation obtained via the Hilbert transform, which can be applied to arbitrary timeseries such as the ones within the simulations and neural datasets used in this study.

Upon first inspection, omitting amplitude dynamics may seem to be an attractive option in neuroimaging studies as the associated dimensional reduction leaves phase as the only dependent variable, hence increasing computational tractability. It is for this reason that phase-only descriptions have allowed for large-scale modelling of dynamic repertoires such as multistability, transience, and criticality, with spatial embedding across the cortical sheet (Breakspear et al., 2010; Ponce-Alvarez et al., 2015). Phase-only models have also been shown to be able to capture patterns of macroscopic neural dynamics consistent with those reported with magnetoencephalography (MEG) and functional magnetic resonance imaging (fMRI) (Cabral et al., 2012; Hellyer et al., 2014), showing similar performance to more involved approaches (Messe et al., 2014).

One of the most widely used phase-only descriptions is the Kuramoto model (Kuramoto, 1975), which was originally developed in the study of chemical phenomena, but has since also been used to explore oscillations in physical, biological, and social systems (Acebron et al., 2005; Strogatz, 2000). The Kuramoto model can capture a range of relatively complicated dynamics, whilst being sufficiently simple to allow for the modelling of a large number of interacting components. The model assumes that different interacting components of a system can be approximated as phase-only oscillators, operating on limit cycles with specific natural frequencies. However, the Kuramoto model pays the same price as any other phase-only description in the treatment of the limit cycle (and its toroidal shell) as a fixed boundary, by leaving the potentially interesting behaviour beyond the limit cycle (Fig. 1B) permanently unexplored.

The potential insights to be gained by generalizing phase-only models to include amplitude dynamics are being increasingly recognised in fields including neuroscience (Ashwin et al., 2016; Park and Ermentrout, 2016), physics (Kurebayashi et al., 2013; Pyragas and Novicenko, 2015), and chemistry (Shirasaka et al., 2017). We set out to derive a dynamic causal model in which oscillators operating close to a supercritical Poincaré–Andronov-Hopf bifurcation (Marsden and McCracken, 1976) – henceforth referred to simply as a Hopf bifurcation – can evolve beyond their limit cycles. This model allows for the inference of coupling strengths in networks of oscillators – something that is only recently being discussed in physics (Marruzzo et al., 2017) and which is investigated here for the first time within the context of Dynamic Causal Modelling (DCM) (Friston et al., 2003) with hypotheses and constraints specific to neural systems.

We begin by making the minimum modifications to the neuronal state equation in DCM required to achieve the broadest possible phase-amplitude description of oscillator dynamics operating close to a Hopf bifurcation. We obtain a full description by measuring the phases and amplitudes of individual oscillators, as well as the angular frequency of the system as a whole (Fig. 1C). Therefore, our work can also be seen as an extension to the Hopf model in (Deco et al., 2017), as we specifically address the issue of input-dependent frequency modulation via the coupling of phase and amplitude. We then use Bayesian methods to determine whether this ability to accommodate phase-amplitude dynamics is advantageous for modelling purposes, given the increased complexity compared with phase-only descriptions.

We then test the models with data that is a) model-generated; b) derived from a simulation of coupled pendula; c) taken from local field potential (LFP) recordings in rodents at different levels of anaesthesia; and d) taken from fMRI recordings in macaque monkeys that are either in an awake or in an anaesthetised state. We find for both the LFP and fMRI datasets, that anaesthetised states are associated with neural dynamics in which oscillator amplitudes play a dominant role. In all four datasets, we use Bayesian model inversion, followed by model reduction, to quantify the relative importance of phase and amplitude effects – allowing for novel data interrogation techniques pertinent to e.g. electrophysiological and fMRI studies of neural function (Daunizeau et al., 2011).

In order to provide a clearer insight into the features of the dynamics captured either by the phase-only or by the phase-amplitude description, we apply our models to four metrics commonly used in neuroimaging studies: the Kuramoto order parameter (Kuramoto, 1975), phase-lag index (Stam et al., 2007), cross-correlation (Stoica and Moses, 2005), and spectral entropy (Zhang et al., 2019). We compare these four metrics as calculated on ground-truth (model-generated) data to the estimates given by the phase-only and by the phase-amplitude models. We find that, with the exception of spectral entropy, the phase-amplitude model outperforms the phase-only model in capturing the metrics.

**Methods**

**The neuronal state equation:** The bilinear form of the neuronal state equation is written as follows, where henceforth we use lowercase, bold, and uppercase fonts to denote scalar, vector, and matrix quantities, respectively:

$$\frac{d\boldsymbol{x}}{dt} = \left(A + \sum_{j}^{M} v_j B^j\right)\boldsymbol{x} + C\boldsymbol{v}, \qquad [1]$$

where $\boldsymbol{x}$ is a vector of size $N \times 1$ that describes the neural activity timecourses in each of the $N$ regions; $A$ is the intrinsic coupling matrix of size $N \times N$ and describes the strength with which the $N$ regions connect to one another in the absence of external inputs; $v_j$ are scalars describing the priors on hidden neuronal causes, in which $j$ indexes the exogenous inputs of which there are a total $M$; $B^j$ are the bilinear coupling matrices, each of which is of size $N \times N$, of which there are a total $M$. Note that the term 'bilinear' derives from the fact that the $B^j$ matrices are defined as second-order derivatives; $C$ is the extrinsic coupling matrix of size $N \times M$ and describes the strength with which the $M$ inputs connect to each of the $N$ regions; and $\boldsymbol{v}$ is a vector of size $M \times 1$ and describes the timecourses of the $M$ exogenous inputs (Friston et al., 2003).

We provide a more detailed explanation of equation [1] in Appendix I.

**The modified neuronal state equation:** here we make the minimum modifications to the complex bilinear form of the neuronal state equation, such that the following three conditions are met: 1) in the limit of weak coupling – i.e. when the amplitudes of the oscillators remain on their respective limit cycles – the model must reduce to the phase-only description given by the Kuramoto model; 2) the intrinsic periodicity of an individual uncoupled oscillator must be maintained in the absence of external perturbations; and 3) phase and amplitude effects must be individually quantifiable upon subsequent model inversion. It is this last condition that necessitates the existence of two separate intrinsic coupling matrices (one for phase and one for amplitude) in the modified state equation - as opposed to the single intrinsic coupling matrix in the original state equation. These two matrices allow the importance of phase and amplitude to be individually quantified upon model inversion and subsequent model reduction.

We begin by re-defining the dependent variable $x$ in [1] to be a complex variable $z$, which we write in polar form as follows:

$$\boldsymbol{x}(t) \; \rightarrow \; \boldsymbol{z}\big(r(t), \theta(t)\big) = \boldsymbol{r}(t) e^{i\boldsymbol{\theta}(t)}, \qquad [2]$$

where $r$ is the amplitude and $\theta$ is the phase.

Note that [2] describes a vector quantity and is therefore to be read as follows:

$$\begin{bmatrix} x_1 \\ x_2 \\ \vdots \\ x_N \end{bmatrix} \rightarrow \begin{bmatrix} z_1 \\ z_2 \\ \vdots \\ z_N \end{bmatrix} = \begin{bmatrix} r_1 e^{i\theta_1} \\ r_2 e^{i\theta_2} \\ \vdots \\ r_N e^{i\theta_N} \end{bmatrix}. \qquad [3]$$

We then re-write [1] using [2], such that:

$$\frac{d\boldsymbol{z}}{dt} = \left( A + \sum_j^M v_j B^j \right) \boldsymbol{z} + C\boldsymbol{v}, \qquad [4]$$

We then define the amplitude $r$ of a given oscillator as the sum of its limit cycle radius $r_{LC}$ and the time-dependent radial distance that its amplitude deviates from the limit cycle (Hale, 1969) due to perturbative effects $r_p$ such that:

$$\boldsymbol{r}(t) = \boldsymbol{r}_{LC} + \boldsymbol{r}_p(t). \qquad [5]$$

Note that $r_{LC}$ is a constant as we are assuming that the oscillators operate close to a Hopf bifurcation, which means that the associated dynamics trace a circular phase portrait.

We then re-write [4] making the two modifications shown in red:

$$\frac{d\mathbf{z}}{dt} = \left(pA + \sum_j^M v_j B^j + Q\right)\mathbf{z} + C\mathbf{v}, \qquad [6]$$

where $p = 1 - \frac{r_{LC}}{|r|}$ is a scalar that represents the fractional radial distance of a given oscillator's amplitude from its limit cycle. Note that $p$ is zero in the limit of weak coupling; in which oscillator dynamics do not evolve beyond their respective limit cycles, i.e. $r = r_{LC}$. Note also that we define $p$ to be a positive quantity as the priors of the $A$ matrix are usually chosen such that its eigenvalues are all real and negative to prevent instabilities; and $Q = idiag\left(\boldsymbol{\Omega} + A_\theta \sum_j^N sin(\boldsymbol{\theta_j} - \boldsymbol{\theta})\right)$ is a matrix of size $N \times N$, the main diagonal of which can be written as a vector $\boldsymbol{\Omega} + A_\theta \sum_j^N sin(\boldsymbol{\theta_j} - \boldsymbol{\theta})$ of size $N \times 1$ that describes the rate of change of phase of each region in the weak coupling limit, where $\boldsymbol{\Omega}$ are the region-specific intrinsic angular frequencies and $A_\theta$ is a phase coupling matrix. Note that this vector describes the Kuramoto model, except that we employ a phase coupling matrix $A_\theta$ in place of the more commonly used normalised global coupling constant.

**Coupled ordinary differential equations of phase and amplitude:** Identifying the real and imaginary components on both sides of equation [6], we obtain the following coupled ordinary differential equations for phase and amplitude:

$$\frac{d\boldsymbol{\theta}}{dt} = \boldsymbol{\Omega} + A_\theta \sum_j^N sin(\boldsymbol{\theta_j} - \boldsymbol{\theta}) - C\mathbf{v} \circ sin\boldsymbol{\theta} \oslash \mathbf{r}, \qquad [7]$$

$$\frac{d\mathbf{r}}{dt} = \left(1 - \frac{r_{LC}}{|r|}\right) A_r \mathbf{r} + \sum_j^M v_j B^j \, \mathbf{r} + C\mathbf{v} \circ cos\boldsymbol{\theta}, \qquad [8]$$

where we use the $\circ$ and $\oslash$ symbols to indicate element-wise vector multiplication and division, respectively; and where we have re-named the intrinsic coupling matrix $A$ in [6] to $A_r$ in [8] as it appears only in the differential equation describing the rate of change of amplitude $r$.

We provide a more detailed explanation of equations [7] and [8] in Appendix II.

**Intrinsic phase and amplitude inter-dependence:** We next modify the $A_\theta$ and $A_r$ matrices in [7] and [8] such that an interdependence between phase and amplitude exists that is independent of perturbative effects in other components of the equations.

We therefore multiply each element of $A_\theta$ in [7] by an amplitude-dependent factor. Specifically, for the coupling strength between the $i^{th}$ and $j^{th}$ regions we multiply $a_{\theta,ij}$ by $e^{-|r_i-r_j|}$. This choice ensures that the unmodified coupling strength $a_{\theta,ij}$ is recovered when $r_i = r_j$. Note that the choice of weighting parameter given by $e^{-|r_i-r_j|}$ assumes that the interaction of interest between two given oscillators depends on the difference between their total amplitudes. However, one might also consider the following alternative weighting factor that incorporates the limit cycle radii of the two given regions: $e^{-|(r_i-r_{LC,i})-(r_j-r_{LC,j})|}$. The latter definition is appropriate when one is interested in the extent to which dynamics are affected specifically by excursions beyond each oscillator's limit cycle. Similarly, we multiply each element of $A_r$ in [8] by a phase-dependent factor. Specifically, for the coupling strength between the $i^{th}$ and $j^{th}$ regions we multiply $a_{r,ij}$ by $cos(\theta_i - \theta_j)$. As with the modification to [7], this choice ensures that the unmodified coupling strength $a_{r,ij}$ is recovered when $\theta_i = \theta_j$.

Using these element-wise weighting factors, we write [7] and [8] in component form to describe the evolution of the $i^{th}$ region, which couples to the $j^{th}$ region and is affected by the $k^{th}$ input:

$$\frac{d\theta_i}{dt} = \Omega_i + \sum_j^N a_{\theta,ij} e^{-|r_i-r_j|} sin(\theta_i - \theta_j) - \frac{c_{ki} v_i sin\theta_i}{r_i}, \qquad [9]$$

$$\frac{dr_i}{dt} = \left(1 - \frac{r_{LC,i}}{|r_i|}\right) a_{r,ij} cos(\theta_i - \theta_j) r_i + v_i B^i r_i + c_{ki} v_i cos\theta_i. \qquad [10]$$

**Limiting case 1 - weak coupling**: here we assess the form of equations [7] and [8] in the limit that coupling between oscillators is weak. In this limit, the amplitudes of the oscillators do not evolve beyond their respective limit cycles and, as we assume that the dynamics operate close to a Hopf bifurcation, we can exploit the associated circular symmetry of the limit cycles in phase space, such that:

$$|r| = r_{LC} \implies \frac{dr}{dt} = 0, \tag{11}$$

which means that [8] is written as follows:

$$0 = \sum_j^M v_j B^j \, r_{LC} + C\boldsymbol{v} \circ cos\boldsymbol{\theta}, \tag{12}$$

which is only true $\forall \, \theta$ if $v = 0$, in which case [7] becomes:

$$\frac{d\boldsymbol{\theta}}{dt} = \boldsymbol{\Omega} + A_\theta \sum_j^N sin(\boldsymbol{\theta_j} - \boldsymbol{\theta}), \tag{13}$$

i.e. we recover the Kuramoto model in the limit of weak coupling.

**Limiting case 2 - single-region intrinsic activity:** here we examine equations [7] and [8] in the limit of a single region that is uncoupled from the network and is unperturbed by external inputs.

Let us consider the $k^{th}$ region as described by [8] and set $v = 0$:

$$\frac{dr_k}{dt} = -\left(1 - \frac{r_{LC,k}}{|r_k|}\right) a_{kk} r_k. \tag{14}$$

where we have defined $a_{kk}$ to be a negative quantity to prevent instabilities.

Given that we are now dealing with scalar quantities and that $r_k > 0$ we can write that $r_k = |r_k|$, which means that [14] becomes separable:

$$\int \frac{dr_k}{r_k - r_{LC,k}} = -a_{kk} \int dt, \tag{15}$$

which has the following solution:

$$r_k(t) = \alpha e^{-a_{kk} t} + r_{LC,k}, \tag{16}$$

where $\alpha$ is a constant of integration.

We see from [16] that, given sufficient time, the amplitude of the isolated region returns to its limit cycle radius:

$$\lim_{t \to \infty} r(t) = r_{LC,k}. \quad [17]$$

Furthermore, from [7] we see that, in the limit of a single region with zero external input, periodicity is maintained as determined by:

$$\frac{d\theta_k}{dt} = \Omega_k, \quad [18]$$

Therefore, [16] and [18] show us that a single region that is uncoupled from the network and is unperturbed by driving inputs will return to its limit cycle whilst oscillating with its intrinsic angular frequency; hence describing a driven, damped harmonic oscillator.

**Coupled stochastic differential equations for phase and amplitude:** all numerical methods used in this paper accommodate noise terms $\omega^{(\theta)}$ to [7] and $\omega^{(r)}$ to [8]:

$$\frac{d\boldsymbol{\theta}}{dt} = \boldsymbol{\Omega} + A_\theta \sum_j^N sin(\boldsymbol{\theta}_j - \boldsymbol{\theta}) - C\boldsymbol{v} \circ sin\boldsymbol{\theta} \oslash \boldsymbol{r} + \boldsymbol{\omega}^{(\theta)}, \quad [19]$$

$$\frac{d\boldsymbol{r}}{dt} = \left(1 - \frac{r_{LC}}{|r|}\right) A_r \boldsymbol{r} + \sum_j^M u_j B^j \boldsymbol{r} + C\boldsymbol{v} \circ cos\boldsymbol{\theta} + \boldsymbol{\omega}^{(r)}, \quad [20]$$

where $\omega^{(\theta)}$ and $\omega^{(r)}$ represent random non-Markovian fluctuations on phase and amplitude, respectively; and random fluctuations to the driving input, $v$, are included via $v = u + \omega^{(v)}$ (Li et al., 2011).

**Simulation of coupled pendula:** A customisable simulation of three coupled pendula is implemented using the physics simulator as part of the Unity3D gaming engine (version 2017.3.1f1). The simulation consists of three pendula that are coupled to one another by spring joints, each of which consists of a bob that is connected by a fixed joint to a rigid body cylinder, which in turn is connected by a hinged joint to a stationary hook. The bob of the first pendulum receives a force that is applied perpendicularly to the hinge joint and follows a Gaussian 'bump' function for the first 5 seconds of each simulation, with the same parameters used for the model-generated data. Each simulation is run for a total of 30 seconds at a sampling rate of 60Hz.

**LFP data:** We assess the importance of phase and amplitude effects under four levels of isoflurane anaesthesia in two-channel LFP recordings from primary and secondary rodent auditory cortices (Moran et al., 2011).

**fMRI data:** The functional MR images are taken from the Nathan Kline Institute Macaque Dataset 1, in which twelve fMRI scans are acquired in an anaesthetized state and twelve in an awake state (Xu et al., 2018).

**fMRI data pre-processing**: We pre-process the data using the FSL image analysis tools as follows: each 4D fMRI scan is registered to a single echo planar image from an earlier session from the same monkey. Focusing on a single monkey in this way simplifies pre-processing as there is no need to perform inter-subject registration. Data is motion corrected, spatially smoothed with a 3mm FWHM Gaussian kernel, and high-pass temporally filtered (100 seconds). Data are then down-sampled into 2x2x2mm voxel space. A data-driven approach is used to define regions for time series extraction. Temporal-concatenation probabilistic independent component analysis (ICA) (Beckmann and Smith, 2004) is then performed across all 24 scans, extracting 20 components (i.e., 20 whole-brain spatial maps). We subsequently perform a spatial multiple linear regression of the 20 components onto the 4D fMRI dataset, resulting in a time course for each component. As expected, many components reflect noise sources (e.g., movement and physiological non-neural signals) - these are excluded, leaving a subset of six components broadly similar to the intrinsic connectivity networks observed in humans. We retain the first two non-noise components (in terms of explained variance) for the purpose of the DCMs presented in this study.

**Bayesian model inversion and model reduction:** We estimate the optimal Bayesian states ([19] and [21]) for parameters (A and C) and hyperparameters (variance of states and parameters) in the Statistical Parametric Mapping (SPM) software using Dynamic Expectation Maximisation (DEM). This inversion routine uses a Laplace approximation of states and parameters (multivariate Gaussians) in generalised coordinates of motion. That is, we estimate not only the rates of change of phase and amplitude ([19] and [21]) but also higher derivatives. We apply 4 embedding dimensions, which accommodate analytic (i.e., smooth) noise processes, unlike the martingale assumptions used in traditional Kalman filters (Roebroeck et al., 2011).

The optimisation scheme employs a DEM algorithm (Friston et al., 2008). DEM is a variational Bayesian scheme with three steps that optimise an approximate posterior over three unknown quantities (i.e., states, parameters and hyper parameters). The D step corresponds to state estimation using variational Bayesian filtering in generalised coordinates, which can be regarded as an instantaneous gradient descent in a moving frame of reference. The E step uses gradient ascent on the negative variational free energy to estimate model parameters and the M step then does the same for the hyperparameters, i.e. precision components of random fluctuations on the states and observation noise (Friston et al., 2008). With reference to equations [19] and [21], we seek to estimate $\theta$ and $r$ via the D step, $A_\theta$, $A_r$ and $C$ via the E step, and finally $\omega^{(\theta)}$ and $\omega^{(r)}$ via the M step. All prior means, precisions and hyperpriors for the log precisions over observation noise are listed in the accompanying code: (https://github.com/allavailablepubliccode/Phase-Amplitude).

Having applied the optimization to the model comprising both phase and amplitude states ([19] and [21]), we then use Bayesian model reduction (Friston and Penny, 2011; Friston et al., 2016) to estimate the evidence for reduced models. Reduced models are specified by setting the prior variance over the off-diagonal parameter values of the phase, amplitude, or both phase and amplitude connectivity matrices to zero. This form of Bayesian model comparison evaluates the variational free energy approximation to log model evidence $logp(y|m)$ for the reduced models and the reduced posterior density over parameters, where the variational free energy combines both accuracy and complexity when scoring models:

$$F = \underbrace{\langle log\,p\,(y|\theta,m)\rangle}_{accuracy} - \underbrace{KL[q(\theta),p(\theta|m)]}_{complexity}, \qquad [21]$$

Here, $logp(y|\theta,m)$ is the log likelihood of the data $y$ conditioned upon model states, parameters and hyperparameters $\theta$, and model structure $m$. In this study we have four model structures: no coupling; phase-only; amplitude-only; and phase-amplitude; $q(\theta)$ is the approximate posterior density over $\theta$; and $p(\theta|m)$ is the prior probability of $\theta$ given $m$, where the Kullback-Leibler divergence ($KL$) term penalises over-complex solutions $q(\theta)$.

In previous work it has been shown that the variational free energy is a better approximation to log model evidence, when compared with the Akaike and Bayesian Information criteria (Penny, 2012). Equation [21] can be rearranged to show that the variational free energy is a lower bound on negative log evidence (known as the ELBO in machine learning) (Winn and Bishop, 2005).

We seek a model in which model evidence is maximised; i.e. provides an accurate explanation for the data in the simplest way possible. For example, we will see that the phase-only model is preferred in some instances – due to the full phase-amplitude model being penalised for complexity. Posterior model probabilities are derived by applying a sigmoidal softmax function $\left(\frac{1}{1+e^{\Sigma F}}\right)$ to the variational free energy bound on log evidence.

**Neuroimaging metrics:** here we test our models on the four commonly used metrics in neuroscience listed below.

Metric 1: the Kuramoto order parameter (KOP) for the $j^{th}$ region is given by:

$$KOP_j = \frac{1}{N}\sum_{j=1}^{N} e^{i\theta_j}. \qquad [22]$$

Metric 2: the cross-correlation (CC) between the $j^{th}$ and $k^{th}$ region is written as follows:

$$CC_{jk} = \int_{-\infty}^{\infty} z_j^*(t)\, z_k(t+\tau)dt, \qquad [23]$$

where $\tau$ is the lag and we use star notation to indicate complex conjugation.

Metric 3: the phase-lag index (PLI) between the $j^{th}$ and $k^{th}$ region is defined as:

$$PLI_{jk} = |\langle sign(\theta_j - \theta_k)\rangle|, \qquad [24]$$

where the angular brackets denote an average across time.

Metric 4: the spectral entropy (SE) for the $j^{th}$ region is given by:

$$SE_j = -\sum_{f=1}^{F} p_j(f) log_2 p_j(f), \qquad [25]$$

where $F$ is the total frequency points; $p_j(f)$ is the probability distribution $p_j(f) = \frac{S_j(f)}{\sum_i S_j(i)}$; $S_j(f) = |Z_j(f)|^2$; and $Z_j(f)$ is the discrete Fourier transform of $z_j(t)$.

In the context of the model-generated data simulation we run the model inversion and subsequent reduction a total of 500 times for both the phase-only and phase-amplitude models. We multiply the external Gaussian input each time by a random number between 0.5 and 2.0, in order to obtain a distribution of estimates of the input (model-generated) timeseries. We then calculate temporal and regional mean values of the above four metrics on a) the ground-truth data given by the model-generated timeseries; b) the estimates of the model-generated timeseries given by the full (phase-amplitude) model; and c) the estimates of the timeseries given by the reduced (phase-only) model following model reduction.

**Results**

**Model-generated data:** To establish construct validity, we generate data from a system of three interacting regions that are coupled in terms of their phase and amplitude in a hierarchical fashion; i.e. one region connects to a subordinate oscillator, which in turn connects to another subordinate oscillator (Fig. 2A). The ensuing timeseries are then subject to generalised filtering under competing models to produce posterior densities of the coupling parameters and the evidence for each model. Because the parameters and models generating the data are known, one can establish the estimability of parameters and the identifiability of competing models.

In detail, we simulate data using equations [19] and [21] using a phase-only and phase-amplitude model (Fig. 2B) with an external driving Gaussian 'bump' function of peristimulus time. The resultant time series (Fig. 2C) are Hilbert transformed into their analytic signals, which subsequently constitute the data feature used for model inversion (data fitting), using the generalised filtering scheme. This procedure optimises the states, parameters and hyperparameters of the model for any given multivariate time series. By assuming fairly precise priors on the amplitude of random fluctuations one can recover the parameters (Fig. 2D & E), their maximum a posteriori (MAP) estimates and their posterior covariation (Fig. 2F) and correlation (Fig. 2G). We then use the approximate marginal log probability (variational free energy) of alternative models (Fig. 2H) for subsequent Bayesian model comparison. In particular, we use Bayesian model reduction to assess the probability for models with 1) no phase or amplitude ($M_1$); 2) phase only ($M_2$); 3) amplitude only ($M_3$); and 4) both phase and amplitude ($M_4$) (Fig 2I).

We compare the estimated parameter values (Fig. 2D & E) to the values used to generate the data (Fig. 2B & E).

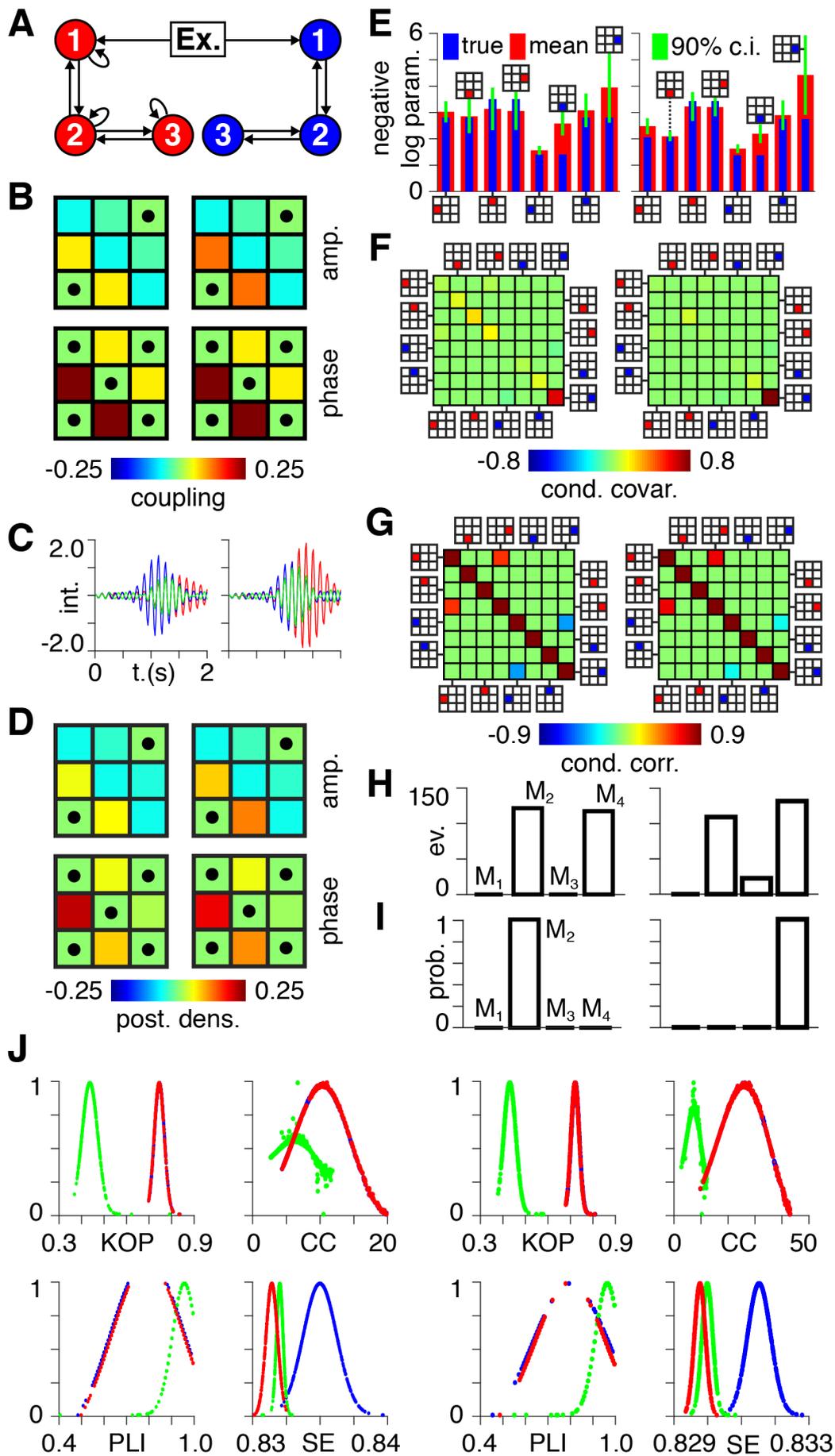

**Fig. 2: synthetic data. (A)** an extrinsic input (Ex.) connects to node 1 of both the phase (blue) and amplitude (red) networks. **(B)** Coupling strengths between nodes of the phase (bottom row) and amplitude (top row) networks for the phase-only (left column) and phase-amplitude (right column) model. Black dots indicate non-existent coupling. **(C)** Signals generated by the phase-only (left) and phase-amplitude (right) models, where the driven node is shown in blue. **(D)** A posteriori estimates of the phase (bottom row) and amplitude (top row) matrices for the phase-only (left column) and phase-amplitude (right column) model. Black dots indicate non-existent coupling. **(E)** All values shown in negative logarithmic space. Mean estimated parameter values (red) vs. true parameter values (blue), together with 90% confidence intervals (green) for the phase-only (left) and phase-amplitude (right) model. The first and last four parameters are the amplitude (red) and phase (blue) coupling elements, respectively, as indicated by the insets. Note that only the first two parameters, i.e. amplitude interaction matrix elements $a_{21}$ and $a_{32}$ vary between the phase-only (left) and phase-amplitude (right) model. **(F)** Conditional co-variance values for phase-only (left) and phase-amplitude (right) models. The first and last four parameters are the amplitude (red) and phase (blue) coupling elements, respectively, as indicated by the insets. **(G)** Conditional correlations for phase-only (left) and phase-amplitude (right) coupling. The first and last four parameters are the amplitude (red) and phase (blue) coupling elements, respectively, as indicated by the insets. **(H)** Approximate log model evidence for phase-only (left) and phase-amplitude (right) scenarios following Bayesian model reduction for models with $M_1$: no phase or amplitude; $M_2$: phase-only; $M_3$: amplitude-only; and $M_4$: phase-amplitude. **(I)** Probabilities derived from the log evidence for the models in (E). **(J)** Kuramoto order parameter (KOP), cross-correlation (CC), phase-lag index (PLI), and spectral entropy (SE) for ground truth model-generated data (red), data recovered via the phase-amplitude model (blue), and data recovered via the phase-only model (green) in the phase-only (left) and phase-amplitude (right) cases.

Prior to model reduction we look for phase-amplitude effects in the estimated coupling matrices and find that non-zero amplitude model effects are captured a posteriori (Fig. 2D & E). Importantly, the inversion recovers stronger amplitude matrix elements for the phase-amplitude model compared with the phase-only model. For the phase estimates, posterior differences are weaker (Fig. 2D & E), in line with the generative models used (Fig. 2B). Figure 2E shows posterior parameter estimates, true parameter values and 90% Bayesian credible intervals for the optimal models, indicating that the model inversion recovers all parameters with reasonable accuracy in both the phase-only (Fig. 2E, left) and phase-amplitude (Fig. 2E, right) models, thereby providing construct validity.

The first two parameters in Figure 2E are the amplitude-only model matrix elements that are varied in order to switch between phase-only and phase-amplitude models (Fig. 2I). The first of these parameters does not fall within the 90% interval for the phase-amplitude model. However, the estimated parameter value remains conservative, in that it lies between the true and the prior value (-3 in logarithmic space). This reflects a well-known effect of shrinkage priors in variational Bayes (Friston et al., 2008). We then evaluate conditional co-variances (Fig. 2F) and conditional correlations (Fig. 2G) between parameters. One can see that the most correlated parameters (indicative of lesser identifiability) are among the amplitude, rather than the phase coupling matrix elements. Bayesian model reduction correctly identifies the structure of the generative model – selecting the correct model (out of four) for data generated with and without amplitude effects (Fig. 2H & I). This means that, under conditions of phase-only coupling, the Bayesian model reduction (which incorporates complexity penalization) correctly identifies the simpler model.

Figure 2J shows that the full phase-amplitude model overestimates spectral entropy but captures (i.e., is statistically indistinguishable from) the ground-truth values of the Kuramoto order parameter, cross-correlation, and phase-lag index. Conversely, the reduced phase-only model fails to capture the Kuramoto order parameter, cross-correlation and phase-lag index, but more closely captures (although is not statistically indistinguishable from) the ground-truth values of the spectral entropy.

**Coupled pendula simulation**: here we use the same procedure as for the model-generated data in Figure 2 to investigate phase and amplitude dynamics in a simulation of three coupled pendula (Fig. 3A & B). Bayesian model reduction is performed following model inversion to calculate posterior parameter estimates and model evidence (Fig. 3C & D) using the same four models as above: no phase or amplitude ($M_1$), phase-only ($M_2$), amplitude-only ($M_3$), and phase-amplitude ($M_4$).

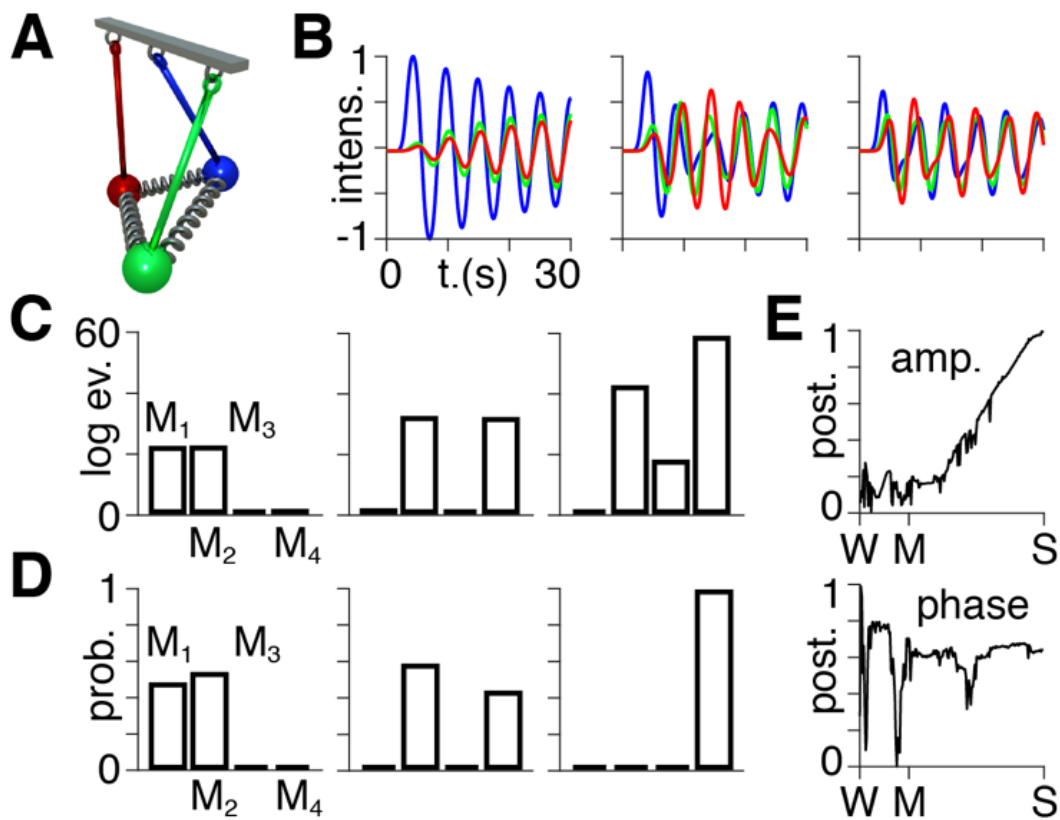

*Fig. 3: Coupled pendula. (A)* Three pendula connected with springs of variable coupling strengths are used to generate data. *(B)* Signals for the pendula (blue = driving pendulum) with weak (left), medium (middle) and strong (right) coupling. *(C)* Approximate lower bound log model evidence given by the variational free energy for weak (left), medium (middle) and strong (right) coupling, for models $M_1$: no phase or amplitude; $M_2$: phase-only; $M_3$: amplitude-only; and $M_4$: phase-amplitude. *(D)* Probabilities derived from the log model evidence for the models in (C). *(E)* Average amplitude (top) and phase (bottom) coupling estimates based on posterior means for a range of coupling strengths with weak (W), medium (M) and strong (S) coupling as shown on the x-axes.

One pendulum is driven by an extrinsic Gaussian input function lasting for 5 seconds and the resultant motion of all three pendula are recorded for a total of 30 seconds at a sampling rate of 60 Hz. The coupling strengths between the pendula are then progressively increased, ranging between weak, medium and strong (Fig. 3B,C &D). We find that the weak coupling scenario is best explained by the 'no phase or amplitude' and 'phase-only' models; the medium coupling by the 'phase-only' and 'phase-amplitude' models; and the strong coupling by the 'phase-amplitude' model. Furthermore, the average posterior connectivity estimates of amplitude-only effects (pooled over elements of the coupling matrix) increases with the coupling strength (r=0.93, p<0.001, Spearman) (Fig. 3E, top). These averages are normalised between zero and unity in the figure, but individual estimates are occasionally negative. No relationship is observed between the average phase coupling matrix elements and spring coupling strength (Fig. 3E, bottom).

**Neuroimaging data**: here we use the same method as for the simulations above to assess the importance of phase and amplitude effects under different levels of anaesthesia in LFP recordings in rodents (Fig. 4A), as well as in fMRI recordings in macaque monkeys (Fig. 4B).

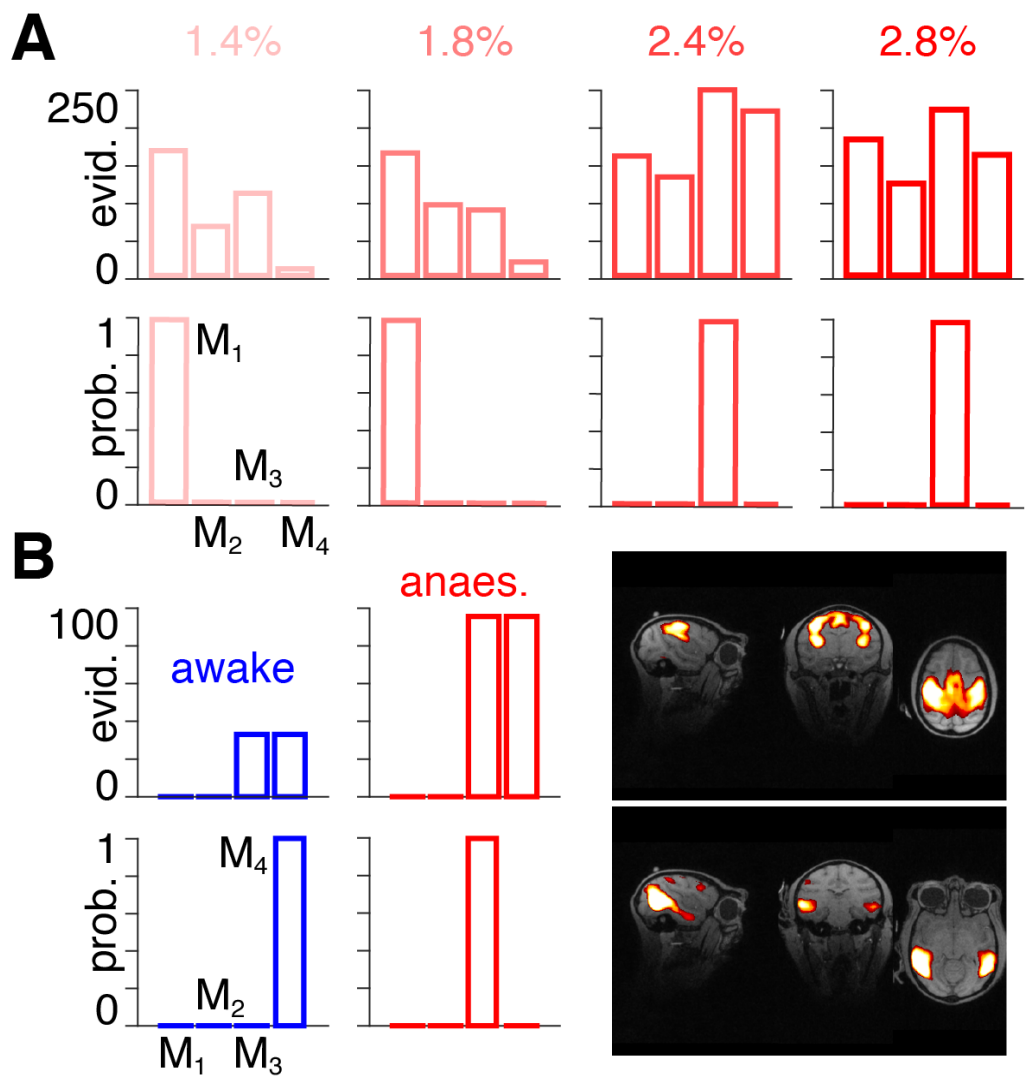

*Fig. 4: Neuroimaging data (A)* LFP recordings in rodents in which four doses of isoflurane are applied intraperitoneally at 1.4, 1.8, 2.4 and 2.8% shown in columns 1, 2, 3, and 4, respectively. Approximate log model evidence given by the *variational* free energy (top row) for models with $M_1$: no phase or amplitude, $M_2$: phase-only, $M_3$: amplitude-only, and $M_4$: phase-amplitude. Probabilities are shown in the bottom row corresponding to the log model evidence values in the top row. *(B)* fMRI recordings in macaque monkeys either in an awake

*(left column) or anaesthetised (right column) state. Approximate log model evidence (all values to be multiplied by $10^3$) given by the variational free energy (top row) for models with $M_1$: no phase or amplitude, $M_2$: phase-only, $M_3$: amplitude-only, and $M_4$: phase-amplitude. Probabilities shown in the bottom row correspond to the log model evidence values in the top row. Regions defined via temporal-concatenation probabilistic ICA are shown on the right, thresholded at z>3, with the first (top) and second (bottom) non-noise components (in terms of explained variance) used in this analysis.*

The model used to explain both these datasets consists of two bi-directionally connected nodes. Bayesian model reduction is performed following inversion to evaluate the evidence for the same four models of coupling: no phase or amplitude ($M_1$), phase-only ($M_2$), amplitude-only ($M_3$), and phase-amplitude ($M_4$).

The results show that the two lowest doses of anaesthesia in the LFP data are best explained by the 'no phase or amplitude' model. On the other hand, the data from the two highest doses of anaesthesia are best explained by the 'amplitude only' model. For the fMRI data the awake state is best explained by the 'phase-amplitude' model and the anaesthetised state by the 'amplitude-only' model. Both the LFP and fMRI results indicate that using a phase-only model in data collected under anaesthesia omits potentially valuable information about the system's dynamics.

**Discussion**

This paper introduces a dynamic causal model that allows for a description of phase-amplitude interactions in a network of oscillators operating close to Hopf bifurcations. We incorporate this model into the DCM framework and establish that it can be used to make inferences about directed phase and amplitude effects over a range of data sources. This provides proof of principle that one can model a broader dynamic repertoire by lifting the weak coupling assumption used by phase-only models. We also demonstrate that phase-amplitude models outperform phase-only models in several scenarios, despite the increased model complexity associated with the inclusion of amplitude as an additional dependent variable. It is important to note that the 'amplitude-only' models can still produce phase-amplitude interactions and also that phase-amplitude models can result in aperiodic signals, in which the term 'phase' is undefined. Furthermore, there is evidence that sinusoids may not be ideally suited to characterise periodic neural signals (Cole and Voytek, 2017). Future work could therefore explore alternative formulations of the model used here with different types of waveforms.

The current approach allows us to explicitly quantify the separate contributions from phase and amplitude in a manner which is computationally tractable and therefore able to be scaled across multiple regions. There are several theoretical accounts of oscillators that describe both phase and amplitude (Hale, 1969; Wedgwood et al., 2013; Wilson and Moehlis, 2016). The modelling approach we use here is guided by the specific aim of quantifying neural oscillations measured with techniques such as LFP and fMRI within the context of DCM – an ability that is relevant within neuroimaging methodologies (Daunizeau et al., 2011).

Four common metrics used in neuroscience are generated from: ground-truth model-generated timeseries; estimates from a phase-amplitude model; and estimates from a phase-only model. This analysis allows for an intuitive handle on the type of dynamics that are captured by the inclusion or exclusion of amplitude as a dependent variable. We find that the phase-amplitude model correctly captures all metrics with the exception of spectral entropy. On the other hand, the phase-only model fails to capture all four metrics, albeit with a better approximation of spectral entropy than the phase-amplitude model. This result demonstrates the need to tailor the model to the specific metrics one is using to interrogate the dataset in question.

We then show that amplitude effects, above and beyond phase effects, are an important feature of: data generated by a simulation of coupled pendula; data acquired at different levels of anaesthesia in LFP recordings in rodents; and fMRI recordings in macaque monkeys either in an awake or in an anaesthetised state. In the LFP data we observe a shift in maximum model evidence from the 'no phase or amplitude' model at low levels of anaesthesia, to the 'amplitude-only' model at high levels of anaesthesia. In the case of the fMRI data, we find that the awake and anaesthetised states are best modelled by the phase-amplitude and amplitude-only models, respectively. Therefore, in both the LFP and fMRI datasets the anaesthetised brain states are associated with an increased importance of oscillator amplitudes. This result is consistent with studies showing that anaesthesia leads to neuronal dynamics that are less complex (Fagerholm et al., 2016) and that operate further from a phase transition (Scott et al., 2014). Furthermore, the increased importance of amplitude compared with phase under anaesthesia is consistent with a bi-stable system that alternates irregularly between stable low and high-activity points (Jercog et al., 2017; Volo et al., 2019), as opposed to a limit cycle that would yield regular behaviour with well-defined phase organisation (Brunel, 2000). Our proposed models may therefore have potential use in determining whether variations in correlation across levels of consciousness; e.g. in anaesthesia (Bettinardi et al., 2015), are due to increased neural participation in oscillations and signal amplitude enhancement, or instead due to increased phase coupling between oscillators.

Models that attempt to explain mechanisms underlying neuronal coupling proffer hypotheses of how features emerge and break down in neuropsychiatric disorders. Using a model such as ours allows for a formal characterisation of how different neuronal dynamics, cognitive tasks or sources of individual variability (e.g., related to pathology) are driven by alterations to phase or amplitude, or a combination of both. The particular model used in this paper is derived from the premise that, by making the minimum required modifications to the DCM neuronal state equation, it is possible to quantify phase and amplitude effects in coupled oscillators operating close to a periodic attractor. Future work with this type of model could provide insight into pathophysiological states that are thought to be associated with a disruption in synchronisation between neural regions, for instance in epilepsy and movement disorders (Buzsaki, 2006; Helfrich et al., 2018; Litvak et al., 2011).

**Appendix I**

We begin by writing the neuronal state equation associated with the $j^{th}$ region, which is uncoupled from the rest of the network and is not affected by exogenous inputs:

$$\frac{dx_j(t)}{dt} = a_{jj} x_j(t), \qquad [26]$$

where $x_j(t)$ is a scalar describing the neural activity in the $j^{th}$ region; and $a_{jj}$ is the intrinsic self-coupling strength of the $j^{th}$ region. Note that such self-coupling strengths are usually given negative priors to prevent instabilities in the dynamics.

Next, we consider the system in [26] when coupled to the $k^{th}$ exogenous input, where we henceforth omit the independent variable for the sake of clarity:

$$\frac{dx_j}{dt} = a_{jj} x_j + c_{kj} v_k, \qquad [27]$$

where $c_{kj}$ is the external coupling strength with which the $k^{th}$ input connects to the $j^{th}$ region; and $v_k$ is a scalar describing the time-dependent $k^{th}$ input.

We now write [27] as a system of differential equations describing the interactions between a total of $N$ regions and $M$ external inputs:

$$\frac{d\boldsymbol{x}}{dt} = A\boldsymbol{x} + C\boldsymbol{v}, \qquad [28]$$

By modifying [28] to accommodate the strength with which the $M$ inputs affects the $N \times N$ connections between the regions we then arrive at the bilinear form of the neuronal state equation introduced in (Friston et al., 2003):

$$\frac{d\boldsymbol{x}}{dt} = \left(A + \sum_{j}^{M} v_j B^j\right)\boldsymbol{x} + C\boldsymbol{v}, \qquad [29]$$

**Appendix II**

We begin by writing [6] in the following form for the sake of clarity:

$$\frac{d\mathbf{z}}{dt} = (X + iY)\mathbf{z} + C\mathbf{v}, \qquad [30]$$

$$X = \left(1 - \frac{r_{LC}}{|r|}\right)A + \sum_{j}^{M} v_j B^j, \qquad [31]$$

$$Y = diag\left(\mathbf{\Omega} + A\sum_{j}^{N} sin(\boldsymbol{\theta_j} - \boldsymbol{\theta})\right). \qquad [32]$$

The dependent variable $\mathbf{z}$ in [30] is complex and can therefore be written in polar form:

$$\mathbf{z}(t) = \mathbf{r}(t)e^{i\boldsymbol{\theta}(t)}, \qquad [33]$$

which we differentiate in time as:

$$\frac{d\mathbf{z}}{dt} = \frac{d\mathbf{r}}{dt}e^{i\boldsymbol{\theta}} + i\frac{d\boldsymbol{\theta}}{dt}\mathbf{r}e^{i\boldsymbol{\theta}}. \qquad [34]$$

Using [33] and [34] we write [30] as:

$$\frac{d\mathbf{r}}{dt}e^{i\boldsymbol{\theta}} + i\frac{d\boldsymbol{\theta}}{dt}\mathbf{r}e^{i\boldsymbol{\theta}} = (X + iY)\mathbf{r}e^{i\boldsymbol{\theta}} + C\mathbf{v}, \qquad [35]$$

which, using Euler's formula: $e^{i\theta} = cos\theta + isin\theta$ can be written as:

$$\left(\frac{d\mathbf{r}}{dt} + i\mathbf{r}\frac{d\boldsymbol{\theta}}{dt}\right)(cos\boldsymbol{\theta} + irsin\boldsymbol{\theta}) = (X + iY)\mathbf{r}(cos\boldsymbol{\theta} + isin\boldsymbol{\theta}) + C\mathbf{v}, \qquad [36]$$

the real components of which satisfy the following:

$$cos\boldsymbol{\theta}\frac{d\mathbf{r}}{dt} - \mathbf{r}sin\boldsymbol{\theta}\frac{d\boldsymbol{\theta}}{dt} - X\mathbf{r}cos\boldsymbol{\theta} + Y\mathbf{r}sin\boldsymbol{\theta} - C\mathbf{v} = 0. \qquad [37]$$

Similarly, the imaginary components of [36] satisfy:

$$sin\boldsymbol{\theta}\frac{d\mathbf{r}}{dt} + \frac{d\boldsymbol{\theta}}{dt}\mathbf{r}cos\boldsymbol{\theta} - Y\mathbf{r}cos\boldsymbol{\theta} - X\mathbf{r}sin\boldsymbol{\theta} = 0. \qquad [38]$$

We can then simultaneously solve [37] and [38] to obtain:

$$\frac{d\boldsymbol{\theta}}{dt} = Y\mathbf{1} - C\mathbf{v} \circ sin\boldsymbol{\theta} \oslash \mathbf{r}, \qquad [39]$$

$$\frac{d\mathbf{r}}{dt} = X\mathbf{r} + C\mathbf{v} \circ cos\boldsymbol{\theta}, \qquad [40]$$

where **1** denotes a unity vector of size $N \times 1$.

Using [31] and [32] we can write [39] and [40] as follows:

$$\frac{d\boldsymbol{\theta}}{dt} = \boldsymbol{\Omega} + A \sum_{j}^{N} sin(\boldsymbol{\theta}_j - \boldsymbol{\theta}) - C\boldsymbol{v} \circ sin\boldsymbol{\theta} \oslash \boldsymbol{r}, \quad [41]$$

$$\frac{d\boldsymbol{r}}{dt} = \left(1 - \frac{r_{LC}}{|r|}\right) A\boldsymbol{r} + \sum_{j}^{M} v_j B^j \, \boldsymbol{r} + C\boldsymbol{v} \circ cos\boldsymbol{\theta}. \quad [42]$$

**Acknowledgements**

E.D.F. and R.L were funded by the MRC (Ref: MR/R005370/1); R.J.M was funded by the Wellcome/EPSRC Centre for Medical Engineering (Ref: WT 203148/Z/16/Z); I.R.V. was funded by the Wellcome Trust (Ref: 103045/Z/13/Z) and the BBSRC (Ref: BB/S008314/1); K.J.F. was funded by a Wellcome Principal Research Fellowship (Ref: 088130/Z/09/Z)

**Author contributions**

All authors designed and performed research, analysed data and wrote the paper.

**Competing interests**

The authors declare no competing interests.